Banner appropriate to article type will appear here in typeset article

# Wall-damped Faraday waves in horizontally oscillating two-layer fluid flows

**Linfeng Piao and Anne Juel**†

Department of Physics & Astronomy and Manchester Centre for Nonlinear Dynamics, The University of Manchester, Oxford Road, Manchester M13 9PL, United Kingdom

We study experimentally the onset of Faraday waves near the endwalls of rectangular vessel containing two stably-stratified fluid layers, subject to horizontal oscillations. These subharmonic waves (SWs) are excited, because the horizontal inertial forcing drives a harmonic propagating wave which displaces the interface in the vertical direction at the endwalls. We find that the onset of SWs is regulated by a balance between capillary and viscous forces, where the rate of damping is set by the Stokes layer thickness at the wall rather than the wavelength of the SWs. We model the onset of SWs with a weakly-damped Mathieu equation and find that the dimensional critical acceleration scales as $v_m^{1/2}\omega^{3/2}$, where $v_m$ is the mean viscosity and $\omega$ is the frequency of forcing, in excellent agreement with the experiment over a wide range of parameters.

## 1. Introduction

The periodic excitation of a fluid interface has long provided a fertile ground for the exploration of fundamental phenomena in fluid dynamics, such as interfacial instabilities, pattern formation, and nonlinear behaviours. It also features in various engineering applications involving mass and heat transfer, such as chemical mixing, solvent extraction, and oil recovery (Rajchenbach & Clamond 2015; Gaponenko *et al.* 2015). Faraday waves are the archetypical response of a fluid interface to periodic forcing (Faraday 1831). These standing wave patterns, which oscillate subharmonically at half the frequency of forcing, are excited when an initially flat interface is subject to vertical vibration. Benjamin & Ursell (1954) demonstrated that in the inviscid limit, they arise from a parametric instability of the Mathieu equation. Viscous effects can play a crucial role in regulating the onset of instability and selecting the wave pattern (Douady 1990). They are often modelled phenomenologically by introducing a heuristic linear damping term into the Mathieu equation. Resulting onset predictions have been extensively validated against experiments and numerical simulations (Kumar & Tuckerman 1994; Christiansen *et al.* 1995; Kumar 1996).

The importance of viscous effects depends on the thickness of Stokes boundary layers, $\delta$, which vary with the vibrational frequency, relative to other lengthscales in the system such as the wavelength, $k^{-1}$, layer depth, $d$, or lateral extent, $L$. The damped Mathieu equation provides satisfactory predictions of the forcing threshold for the onset of Faraday waves in the limit of deep liquid layers, $kd \gg 1$, and weak viscous effects in the bulk, $k\delta \ll 1$, which corresponds to $\gamma/\omega \ll 1$, where $\gamma$ is the damping rate and $\omega$ the forcing frequency. It has also been modified to model dissipative effects due to the Stokes boundary layers (wall damping) and the moving contact line (Christiansen *et al.* 1995). Wall damping becomes particularly significant in shallow layers, $kd \ll 1$, and layers of small lateral extent, $kL \ll 1$ (Douady 1990; Tipton & Mullin 2004), because $\delta$ becomes an important determinant of interface dynamics (Christiansen *et al.* 1995). Specifically, Lioubashevski (1997) showed that for a thin liquid layer where $k\delta \sim O(10^0)$, the damping rate is $\gamma \sim \omega$. Tipton & Mullin (2004) studied Faraday waves at the interface between two immiscible fluids, but most studies have focused on a single fluid layer with a free surface. In this paper, we consider a two-liquid system and show that wall-damped Faraday waves with $k\delta \sim O(10^{-1})$ can arise in a large container with deep layers ($kL \gg 1$ and $kd \gg 1$), when the periodic forcing is applied in the horizontal direction.

† Email address for correspondence: anne.juel@manchester.ac.uk

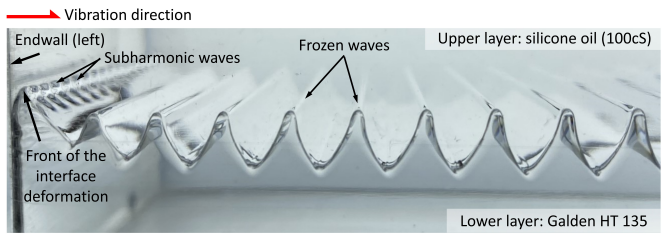

Figure 1: A snapshot of interfacial instabilities occurring on the interface between two immiscible fluids subject to the horizontal forcing with frequency $f = 50$ Hz and amplitude $A = 1.67$ mm.

When a sealed vessel containing two stably stratified layers of immiscible liquids with different densities oscillates horizontally, the fluid layers are differentially accelerated into a counterflow due to the confinement of the endwalls. Away from the endwalls, oscillatory shear of the interface can drive a Kelvin–Helmholtz-type instability above a critical acceleration, which results in the formation of an array of gravity-capillary waves along the direction of oscillation. An example is shown in figure 1 of these 'frozen waves', which appear static in the co-moving frame (Talib *et al.* 2007). We show that subharmonic standing waves (SWs) can also be excited near the endwalls, as shown in figure 1. This is because the inertial counterflow redirects horizontal forcing into vertical oscillation of a localised interfacial front which loses stability beyond a critical value of forcing. When these waves reach sufficient amplitude, they can periodically shed droplets in a controllable manner and may thus offer a useful means of generating bespoke emulsions. However, to the best of our knowledge, the conditions for onset of these SWs have not been established, despite extensive investigation of horizontally oscillating two-layer flows (Jalikop & Juel 2009; Gaponenko *et al.* 2015; Sánchez & Shevtsova 2019). Porter *et al.* (2012) experimentally observed a staggered subharmonic wave pattern at the surface of a viscous liquid in a large horizontally vibrating container. The wave amplitude decays away from the outer rim but the wave pattern extends across the entire container. Using a damped Mathieu equation model with a linear bulk damping term and spatially inhomogeneous parametric forcing, Porter *et al.* (2012) numerically reproduced similar patterns to those observed in their experiments, but they could not satisfactorily predict critical parameter values to match experimental onset measurements, with predicted values consistently below those measured experimentally. Subsequent theoretical studies have elaborated on the localising effect of this spatially inhomogeneous parametric forcing (Tinao *et al.* 2014; Perez-Gracia *et al.* 2014). In this paper, we show that for waves localised close enough to the endwalls so that $k\delta \sim O(10^{-1})$, a Mathieu equation model with an appropriate choice of wall damping can predict the onset of the SWs observed.

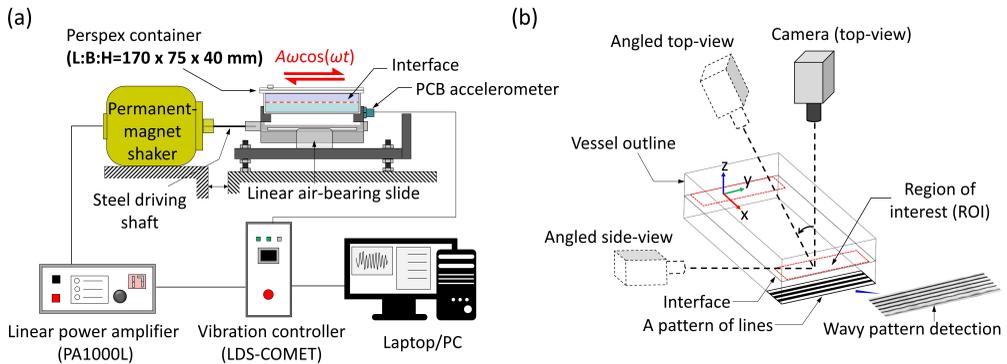

Figure 2: (a) Schematic diagram of the experimental set-up. (b) Visualization set-up for interfacial dynamics near the endwall regions and wavy pattern detection method.

The paper is organised as follows. We present the experimental set-up in §2 and discuss experimental observations of wall effects in §3.1, including how the localisation of the SWs near the endwalls depends on viscosity and vibrational parameters. In §3.2 we show that the critical forcing acceleration at which SWs appear depends on the value of the mean viscosity of the layers. In §3.3, we propose a simple model of the onset of the SWs based on scaling analysis of the Mathieu equation which closely matched our experimental observations, indicating that the onset of the SWs is regulated by viscous dissipation in the Stokes wall layer. Concluding remarks and outlook are given in §4.

## 2. Experimental set-up

A schematic diagram of the experiment is shown in figure 2. A rectangular Perspex container with inner dimensions of length × width × height = 170 mm × 75 mm × 40 mm was filled with equal volumes of immiscible fluids. At rest, the fluids adopted a two-layer stably stratified configuration where they were separated by a flat, horizontal interface. The container was rigidly mounted on a horizontal air-bearing slide (Nelson Air) and connected by a thin steel rod to a permanent magnet shaker (LDS-V450) powered by a linear amplifier (LDS, PA 1000L) and controlled using a vibration controller (LDS-COMET USB). The shaker imposed horizontal harmonic oscillations of the container with prescribed velocity $A\omega \cos(\omega t)$, where $A$ is the amplitude and $\omega = 2\pi f$ the angular frequency. We performed experiments under different vibrational conditions with forcing frequencies in the range $20\,\text{Hz} \leqslant f \leqslant 60\,\text{Hz}$ and forcing amplitudes up to $A = 3.00$ mm. The harmonic content of the motion of the container, measured using an accelerometer (PCB Piezotronics, model 353B43), is less than 0.1% over the entire frequency range. We refer to our previous studies of the frozen-wave instability (Talib *et al.* 2007; Jalikop & Juel 2009) for a detailed description of the shaker system.

We used silicone oil (polydimethylsiloxane fluids, Basildon Chemicals Ltd) and a perfluorinated polyether (Galden® HT fluids, Solvay) for the upper and lower fluid layers, respectively. Table 1 lists the density $\rho$ and kinematic viscosity $\nu$ of the four different grades of silicone oil (SO10–SO100) and two perfluorinated fluids (HT135 and HT270) used. The interfacial tensions $\sigma$ between each Galden fluid and silicone oil at $21 \pm 1$°C were $6.8 \pm 0.5$ mN/m and $8.5 \pm 0.5$ mN/m, respectively, and did not vary measurably with silicone oil grade. The large density difference between the fluid layers promotes significant differential velocities under horizontal acceleration and their low interfacial tension ensures that at rest, the interface is not measurably distorted near the walls (Jalikop & Juel 2009). The depth of each fluid layer was $d = 20$ mm which is more than an order of magnitude larger than the capillary length $l_{ca} = [\sigma/(g\Delta\rho)]^{1/2} \simeq 1.0$ mm, where $\Delta\rho = \rho_l - \rho_u$ is the density difference between the fluids and $g$ the gravitational acceleration. Two high-speed cameras (Photron FASTCAM mini AX100 and PCO.1200 hs) were used to capture interfacial phenomena. The first camera recorded two different views, an angled top view and an angled side view (see figure 2(b)), with respective resolutions of 39.4 and 50.4 pixels/mm and a minimum rate of 4000 frames per second. The second camera was used to record top views at 500 frames per second with a resolution of 64.1 pixels/mm. The fluid interface was lit uniformly by two LED panels, a primary vertical panel behind the container and a secondary horizontal panel above the container. We visualized interfacial deformation in the region of interest (ROI) indicated by a red rectangle near the right endwall in figure 2(b). A pattern comprising 12 black lines parallel to the endwalls, with line thickness 0.2 mm and interline spacing of 1 mm, was positioned under the transparent bottom boundary of the container. The visualisation in top-view of the image of this line pattern, refracted by the fluid interface, enabled the detection of small interface deformations along the $y$-direction down to 0.1 mm, by measuring the distortion of the line. Periodic deformations were detected upon onset of instability and their wavenumber $k$ was determined by $k = 2\pi N/(W/2)$, where the mode number $N$ is defined as the number of wavelengths spanning the half-width $(W/2)$ of the container. To determine $N$, we tracked the edge of the deformed line along the y-direction using MATLAB's Canny algorithm and then analysed it with fast Fourier transform.

| Upper layer | $\nu_u(10^{-6}$ m$^2$/s) | $\rho_u$ (kg/m$^3$) | Lower layer | $\nu_l(10^{-6}$ m$^2$/s) | $\rho_l$ (kg/m$^3$) |
|---|---|---|---|---|---|
| Silicone oil (10 cS) | 10.3 | 935 | Galden HT 135 | 1.12[b] | 1752 |
| Silicone oil (20 cS) | 21.7 | 950 | Galden HT 270 | 11.7[a] | 1856[a] |
| Silicone oil (50 cS) | 54.8 | 961 | | | |
| Silicone oil (100 cS) | 113.7 | 961 | | | |

Table 1: Physical properties of the liquids used in the experiments. The viscosities of silicone oils were measured at 21 ± 1°C using a Kinexus rheometer. [a]From manufacturer's data at 25°C. [b]Measured by Jalikop & Juel (2009) at 21±1°C.

## 3. Results and discussion

### 3.1. *Subharmonic waves confined near the endwalls*

We find that subharmonic waves (SWs) can be excited in the close vicinity of the endwalls of the container depending on the amplitude and frequency of the imposed horizontal forcing, as illustrated in figure 1. The differential inertial forcing of the two fluid layers drives synchronous (harmonic) propagating waves which decay away from the endwalls due to viscous dissipation (Perez-Gracia *et al.* 2014; Sánchez & Porter 2019). Figure 3(b) shows side-view snapshots of this harmonic wave at the right hand-side wall for the HT135-SO50 fluid pair with $f = 35$ Hz and $A = 1.30$ mm, which is below the threshold of onset of SWs. The four snapshots were taken every quarter of a period of oscillation $T$, with corresponding displacements of the container shown in figure 3(a). The images indicate that fluid located within three capillary lengths $(3l_{ca})$ of the wall rises and recedes harmonically with the horizontal oscillation of the container, resulting in the local vertical displacement of the interface (Sánchez & Shevtsova 2019). The interface adjacent to the right hand-side wall is displaced downwards while the container moves to the right, because the denser fluid in the lower layer flows away from the endwall due to inertia. The minimum height of the interface is reached when the container reaches its far-right position (+0.25T). The subsequent change in the direction of motion of the container leads to the development of an opposite counterflow where the lower layer now flows towards the endwall. This in turn drives the upward displacement of the interface at the endwall (+0.5T) up to a maximum height reached at the maximum container displacement (+0.75T). Note that the upward displacement of the interface is larger than its downward displacement because of the reduced viscosity of the lower layer. We observed this harmonic wave which drives interface deformation at the endwalls for all values of the forcing acceleration, $a = A\omega^2$ and fluid pairs investigated. However, the distance from the endwalls over which interfacial deformation was observed, which we refer to as the harmonic wave field, increased from the minimum value of $3l_{\text{ca}}$ as the mean viscosity in the system was reduced. This is because the spread of the wave field is set by the relative magnitude of viscous and capillary effects (Perez-Gracia *et al.* 2014). In figure 3(b), the height of the harmonic wave front is uniform across the width of the channel ($W \approx 75l_{\text{ca}}$). However, an increase in the amplitude of forcing to $A = 1.45$ mm destabilises this flat front into a standing wave pattern in the y-direction as shown in figure 3(c). Figure 3(d) shows the evolution of the standing wave over two periods of oscillation in angled top view for $f = 40$ Hz and $A = 1.30$ mm for HT135-SO50. The two cycles of oscillation shown in figure 3(d) are superposed in figure 3(e) so that each snapshot now combines images one period apart, with the interface from the second cycle highlighted with a red dotted line; see , e.g., the images at the times $t^*$ and $t^* + T$ in figure 3(e). We find a phase shift of 180° between the superposed wave patterns, which indicates that the standing wave oscillates at half of the forcing frequency (i.e., $\omega = 2\omega_0$) and is therefore subharmonic.

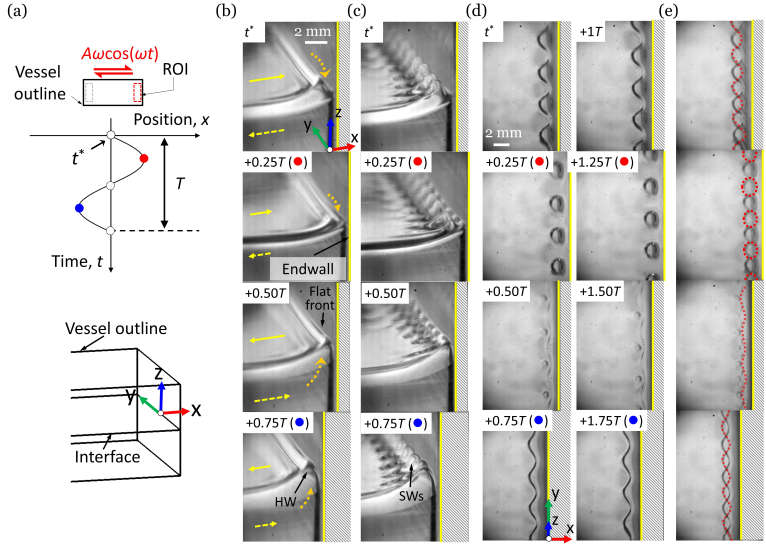

Figure 3: (a) Schematic diagrams showing the *x*-position of the centre of the container over one period (top), and the ROI (bottom). (a-d) Snapshot images of the ROI in different views (see figure 2(b)) for the HT135-SO50 fluid pair, with the *x* displacement of the endwall (vertical yellow line) indicating the instantaneous displacement of the vessel: (b) Harmonic wave front of uniform height along *y*, visualised in angled side view ($f = 35$ Hz, $A = 1.30$ mm). (c) Development of subharmonic waves (SWs) on the harmonic wave front in angled side view ($f = 35$ Hz, $A = 1.45$ mm). (d) Experimental snapshots of the SWs visualized in angled top view ($f = 40$ Hz, $A = 1.30$ mm). (e) Superimposed 180° out-of-phase images for patterns observed in (d).

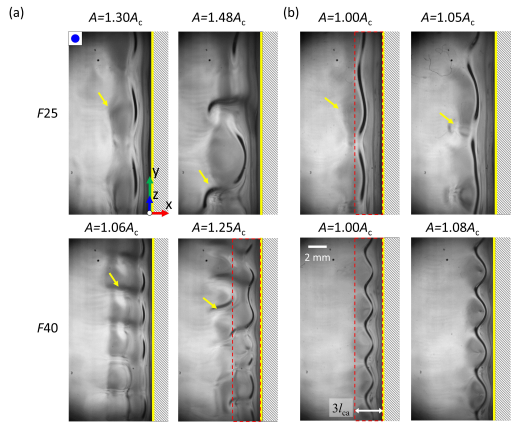

Figure 4: Subharmonic wave patterns near the endwalls for fluid pairs (a) HT135-SO10 and (b) HT135-SO50, where images are captured in angled top view (see figure 2(b)) with the container at its leftmost position (blue circle as in figure 3), for different forcing frequencies: $f$=25 Hz ($F$25) and $f$=40 Hz ($F$40). The coordinate system shown was defined in figure 3(a). The yellow line indicates the position of the endwall and the yellow arrow points to the appearance of SWs outside the near-endwall region ($\sim 3l_{ca}$, as indicated by the red boxes).

The distance over which the harmonic wave field decays away from the endwall determined the spread of the SWs in the *x*-direction. Figure 4 compares SWs excited in experiments with the fluid pairs HT135-SO10 (figure 4(a)) and HT135-SO50 (figure 4(b)). The upper and lower rows show experiments performed with $f$=25 Hz ($F$25) and 40 Hz ($F$40), respectively, and amplitudes of forcing are given relative to the onset value, $A_c$, of the SWs. The yellow arrows highlight subharmonic waves which extend beyond the near-end wall region where the harmonic wave crest oscillates vertically ($\sim 3l_{ca}$). SWs formed in the HT135-SO10 fluid pair with lower mean viscosity tend to extend beyond the near-wall region regardless of the forcing frequency; in contrast, SWs formed in the HT135-SO50 fluid pair with higher mean viscosity localise in the close vicinity of the endwall for the higher forcing frequency. The harmonic forcing is only vertical in the near-endwall region, beyond which the harmonic wave front is subject to a combination of vertical and horizontal forcing, which renders the SWs spatially inhomogeneous (Porter *et al.* 2012; Tinao *et al.* 2014). For less viscous fluid layers or low frequencies of forcing, wave interaction upon a small increase in the forcing amplitude beyond the onset of the SWs can lead to irregular patterns, as illustrated in figure 4. Hence the nature of the SWs observed experimentally differs depending on vibrational parameters and mean viscosity. However, at onset, they always arise next to the endwall where vertical forcing is strongest.

### 3.2. *Effect of fluid properties and vibrational parameters*

Figure 5(a) shows the threshold forcing acceleration $a_c = A_c \omega^2$ at which SWs appear as a function of the forcing frequency for the five fluid pairs introduced in §2. These onset measurements were performed by imposing the frequency of forcing and gradually increasing the amplitude of forcing in increments of 0.01 mm until the first appearance of the instability at the threshold amplitude $A_c$, taking care to let transients decay for at least 600 cycles of oscillation after each increment. We find that for each fluid pair, the threshold acceleration increases monotonically with increasing forcing frequency. Remarkably, the onset curves are stacked in order of increasing mean kinematic viscosity, $\nu_m = (\rho_u \nu_u + \rho_l \nu_l)/(\rho_u + \rho_l)$, with the lowest curve corresponding to the lowest value of $\nu_m$. Note that the viscosity ratio $\nu_u/\nu_l$, which governs the onset of the frozen wave instability (Talib *et al.* 2007), does not increase monotonically with $\nu_m$ for the experiments shown in figure 5(a). This is because of the red data which has the lowest viscosity ratio, due to the tenfold increase of the lower layer viscosity in these experiments compared to the other data shown with black symbols. In fact, the results of figure 5(a) bear no resemblance to the onset of the frozen wave instability, whose threshold acceleration decreases monotonically as a function of frequency, and whose onset curves are stacked in order of decreasing $\nu_u/\nu_l$. Figure 5(a) also indicates that the datasets do not remain similar as $\nu_m$ is increased, because of increasingly different threshold acceleration values at high frequency. This means that a rescaling in terms of only the mean viscosity is not sufficient to collapse the experimental data onto a mastercurve, and we refer to §3.3 for a more detailed scaling analysis.

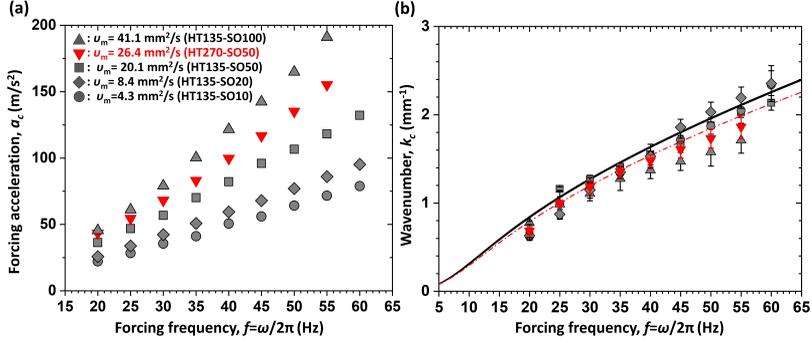

Figure 5: Critical acceleration (a) and critical wavenumber (b) as a function of forcing frequency indicating the onset of SWs in the experiment for five fluid pairs with different mean kinematic viscosity, $\nu_m$, represented by different symbols. Experimental measurements in (b) are compared to the Faraday wave dispersion relation (solid lines) for each fluid pair. The red line sits slightly below the black lines because the density of HT270 is approximately 5% larger than that of HT135 (see table 1).

The dispersion relation of the SWs is shown in figure 5(b), where the wavenumber ($k_c$) measured at onset is plotted as a function of frequency for the five fluid pairs investigated. The symbols used to indicate different values of $\nu_m$ are the same as in figure 5(a). Each data point represents the average of several separate experiments, and the error bar indicates the standard deviation of this value. The lines show the theoretical prediction for Faraday waves at the interface between two infinite-depth fluid layers, which is given by the gravity-capillary wave dispersion relation, $(1/2\omega)^2 = \omega_0^2 = (\sigma k^3 + \Delta\rho g k)/(\rho_l + \rho_u)$, where $\omega_0$ is the natural angular frequency (Kumar & Tuckerman 1994; Rajchenbach & Clamond 2015). The experimental dependence of wavelength on frequency is captured satisfactorily by this dispersion relation. The agreement is within error bars for the lowest values of $\nu_m$, but for $\nu_m > 20.1$ mm$^2$/s, the experimental data sits marginally below the theoretical lines. This discrepancy is associated with the increase in the rate of viscous dissipation with forcing frequency and mean viscosity, which has been previously highlighted in studies of Faraday waves (Edwards & Fauve 1994; Bechhoefer *et al.* 1995).

### 3.3. *Scaling analysis and physical interpretation*

Measured wavelengths on the order of the capillary length, and the increase of the critical acceleration for the onset of SWs with increasing $\nu_m$ discussed in §3.2, suggest that the onset of SWs is governed by the relative magnitude of capillary and viscous forces. Hence, we follow Goodridge *et al.* (1997) in defining a capillary-viscous length, $l_{cv} = \nu_m^2/(\sigma/\Delta\rho)$, and a capillary-viscous frequency, $\Omega_{cv} = (\sigma/\Delta\rho)^2/(\nu_m^3)$, and use them to scale the critical forcing acceleration $a_c^* = a_c/(l_{cv}\Omega_{cv}^2)$ and the forcing frequency, $\omega^* = \omega/\Omega_{cv}$. Figure 6(a) shows that the scaled critical acceleration data previously shown in figure 5(a) approximately collapses onto a master curve proportional to $\omega^{*1.46\pm0.01}$, which we obtained by linear regression of the experimental data (blue dashed line). The scaled critical wavenumber, $k_c^* = k_c l_{cv}$, is plotted as a function of $\omega^*$ in the inset of figure 6(a), where the solid lines correspond to the capillary-gravity wave dispersion relation shown in figure 5(b). The scaled dispersion relation is shifted to larger frequencies as $\nu_m$ is increased. The upper parts of each of these curves which correspond to sufficiently large values of $\omega^*$ approximately collapse onto a single line indicating capillary-viscous wavenumber selection. We find that each set of experimental data aligns onto the appropriate dispersion relation curve close to this line, which further supports our choice of scaling.

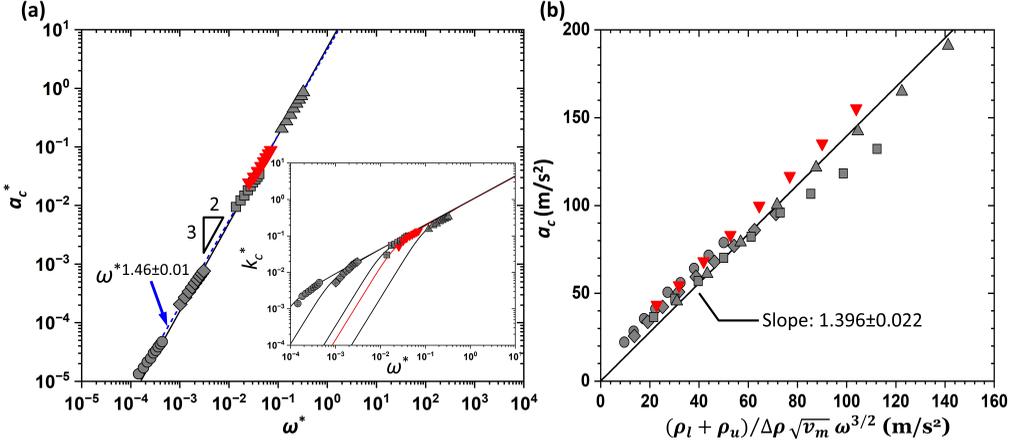

Figure 6: (a) Log-log plot of non-dimensional acceleration versus non-dimensional frequency for different fluid pairs. A least-square fit yields $a_c^* \sim \omega^{*1.46\pm0.01}$ (dashed line) and the solid line indicates $a_c^*\omega^{*3/2}$ predicted by equation (3.2) (solid line). Inset: Nondimensional wavenumber versus $\omega^*$. The solid lines indicate the Faraday wave dispersion relation on the capillary-viscous scale. (b) Experimental critical acceleration $a_c$ plotted against the prediction of equation (3.2).

To predict the critical value of scaled acceleration for the onset of SWs within the limit of weak viscous effects (Kumar 2000), we use the Mathieu equation with linear viscous damping as a simple model describing a parametrically forced interface separating infinite-depth layers (Kumar & Tuckerman 1994; Porter *et al.* 2013). The dimensional critical acceleration is given by

$$a_c = 2\gamma(\rho_l + \rho_u)\omega/(\Delta\rho k_c). \qquad (3.1)$$

Following Landau & Lifshitz (1987), the damping rate is defined as $\gamma = \overline{\dot{E}}/(2\overline{E})$, where $\overline{\dot{E}}$ and $\overline{E}$ denote the time-averaged rate of viscous dissipation and the time-averaged mechanical energy, respectively. Kumar & Tuckerman (1994) show that for two fluid layers of the same depth, $d$, the time-averaged rate of dissipation and total mechanical energy can be estimated through volume integrals as $\overline{\dot{E}} = -2(\rho_l \nu_l \int (\nabla \cdot \overline{u}_l)^2 dV + \rho_u \nu_u \int (\nabla \cdot \overline{u}_u)^2 dV)$ and $\overline{E} = \rho_l \int \overline{u}_l^2 dV + \rho_u \int \overline{u}_u^2 dV$, where $\overline{u}_l$, $\overline{u}_u$ are the time-averaged fluid velocities that can be characterised by $u = A(\omega/2)$ (Bechhoefer *et al.* 1995). Using the onset wavelength of the Faraday wave pattern, $\lambda_c = 2\pi/k_c$ as the lengthscale of damping in the limit of deep fluid layers where $kd \gg 1$, the damping rate is evaluated in a volume $\lambda_c^2 W$ as
$$\gamma \sim \nu_m \frac{(u/\lambda_c)^2(W\lambda_c^2)}{(u)^2(W\lambda_c^2)} = \nu_m k_c^2 \,.$$

As shown in §3.1, the SWs are adjacent to the endwalls and thus, the damping rate over $\lambda_c$ is dominated by the Stokes boundary layers of thickness $\delta \equiv (2\nu_m/\omega)^{1/2}$ at these walls. In our experiments, we measured $k_c\delta \approx O(10^{-1})$. The contribution to $\bar{E}$ from the Stokes layer is a factor $(k_c\delta)^{-1}$ larger than that in the bulk whose contribution we neglect. Hence, the end-wall damping rate in a volume $\lambda_c^2 W$ can be evaluated as $\gamma \sim \nu_m \frac{(u/\delta)^2(W\lambda_c\delta)}{(u)^2(W\lambda_c^2)} = \omega\delta/\lambda_c$ (Milner 1991; Christiansen *et al.* 1995), which differs from the bulk damping rate in both limits of weak viscous effects and deep layers (Kumar 1996) and thin layers (Lioubashevski 1997). Using this expression, we can recast equation (3.1) as

$$a_c \sim \frac{(\rho_l + \rho_u)}{\Delta\rho}\sqrt{\nu_m}\omega^{3/2}. \tag{3.2}$$

The frequency dependence in $\omega^{3/2}$ (solid line) is in close agreement with the experimental data in figure 6(a) which has a power exponent of $1.46 \pm 0.01$ (dashed line). We attribute the slight discrepancy with the experiments to imperfect collapse of the datasets associated with the two lowest values of $\nu_m$. The expression (3.2) also indicates dependence of the critical forcing acceleration on the square root of the mean viscosity. We test the theoretical prediction given by equation (3.2) by plotting it against the experimentally measured values of $a_c$ in figure 6(b), using the same symbols for different values of $\nu_m$ as in figure 6(a). Note that we are not able to vary the density difference between the fluids significantly. The experimental data collapses satisfactorily on a straight line and a least-square proportional fit yields a slope of $1.396 \pm 0.022$, which depends on the prefactor of the damping rate. This result indicates that the SWs excited near the endwalls of our horizontally vibrated container are driven through a Faraday instability dominated by a wall-damping mechanism.

## 4. Concluding remarks

We have shown that the horizontal excitation of a vessel containing two superposed immiscible liquid layers can generate Faraday wave instabilities along the walls of the container which are perpendicular to the direction of forcing. This is because the horizontal forcing drives a propagating wave which harmonically displaces the interface in the vertical direction in the vicinity of the endwalls and the uniform wavefront that results becomes unstable to subharmonic standing waves at a critical acceleration. In contrast with the well-studied Kelvin–Helmholtz instability which arises in the central part of the container as frozen waves, the onset of these wall-damped subharmonic waves depends on the mean viscosity rather than the viscosity ratio of the liquids. We find that despite thin Stokes boundary layers on the endwalls so that $k_c\delta \sim O(10^{-1})$, the onset of Faraday waves relies on a balance between capillary and viscous forces and can be modelled by a weakly-damped Mathieu equation. This is because the rate of damping is dominated by the dissipation in the Stokes layer. Scaling analysis of the Mathieu equation yields a dependence of the dimensional critical acceleration on $\nu_m^{1/2}\omega^{3/2}$ in excellent agreement with experiments.

Upon further increase of the horizontal oscillatory forcing of the container beyond the onset of SWs, we observe pinch-off of droplets from the tips of the SWs. By selecting appropriate combinations of the vibrational parameters and fluid viscosities we can produce monodisperse droplets at a rate proportional to the vibration frequency. This offers the prospect, currently under investigation, of a new route to the controlled generation of bespoke emulsions, underpinned by the mechanisms uncovered in this paper.

**Acknowledgements:** This work was supported by a Horizon Europe Guarantee MSCA Postdoctoral Fellowship through EPSRC (EP/X023176/1).

**Declaration of Interests:** The authors report no conflict of interest.

REFERENCES

BECHHOEFER, J., EGO, V., MANNEVILLE, S. & JOHNSON, B. 1995 An experimental study of the onset of parametrically pumped surface waves in viscous fluids. *J. Fluid Mech.* **288**, 325–350.

BENJAMIN, T. B. & URSELL, F. J. 1954 The stability of the plane free surface of a liquid in vertical periodic motion. *Proc. R. Soc. Lond. A* **225**, 505–515.

CHRISTIANSEN, B., ALSTRØM, P. & LEVINSEN, M. T. 1995 Dissipation and ordering in capillary waves at high aspect ratios. *J. Fluid Mech.* **291**, 323–341.

DOUADY, S. 1990 Experimental study of the faraday instability. *J. Fluid Mech.* **221**, 383–409.

EDWARDS, W. S. & FAUVE, S. 1994 Patterns and quasi-patterns in the faraday experiment. *J. Fluid Mech.* **278**, 123–148.

FARADAY, M. 1831 Xvii. on a peculiar class of acoustical figures; and on certain forms assumed by groups of particles upon vibrating elastic surfaces. *Phil. Trans. R. Soc. Lond.* **121**, 299–340.

GAPONENKO, Y. A., TORREGROSA, M., YASNOU, V., MIALDUN, A. & SHEVTSOVA, V. 2015 Dynamics of the interface between miscible liquids subjected to horizontal vibration. *J. Fluid Mech.* **784**, 342–372.

GOODRIDGE, C. L., SHI, W. T., HENTSCHEL, H. G. E. & LATHROP, D. P. 1997 Viscous effects in droplet-ejecting capillary waves. *Phys. Rev. E* **56**, 472–475.

JALIKOP, S. V. & JUEL, A. 2009 Steep capillary-gravity waves in oscillatory shear-driven flows. *J. Fluid Mech.* **640**, 131–150.

KUMAR, K. 1996 Linear theory of faraday instability in viscous liquids. *Proc. R. Soc. A* **452**, 1113–1126.

KUMAR, K. 2000 Mechanism for the faraday instability in viscous liquids. *Phys. Rev. E* **62**, 1416.

KUMAR, K. & TUCKERMAN, L. S. 1994 Parametric instability of the interface between two fluids. *J. Fluid Mech.* **279**, 49–68.

LANDAU, L. D. & LIFSHITZ, E. M. 1987 In *Fluid mechanics, 2nd edn.*. Pergamon.

LIOUBASHEVSKI, O.AND FINEBERG, J.AND TUCKERMAN L. S. 1997 Scaling of the transition to parametrically driven surface waves in highly dissipative systems. *Phys. Rev. E* **55**, R3832.

MILNER, S. T. 1991 Square patterns and secondary instabilities in driven capillary waves. *J. Fluid Mech.* **225**, 81–100.

PEREZ-GRACIA, J. M., PORTER, J., VARAS, F. & VEGA, J. M. 2014 Subharmonic capillary–gravity waves in large containers subject to horizontal vibrations. *J. Fluid Mech.* **739**, 196–228.

PORTER, J, TINAO, I, LAVERÓN-SIMAVILLA, A & LOPEZ, C A 2012 Pattern selection in a horizontally vibrated container. *Fluid Dyn. Res.* **44**, 065501.

PORTER, J., TINAO, I., LAVERÓN-SIMAVILLA, A. & RODRÍGUEZ, J. 2013 Onset patterns in a simple model of localized parametric forcing. *Phys. Rev. E* **88**, 042913.

RAJCHENBACH, J. & CLAMOND, D. 2015 Faraday waves: their dispersion relation, nature of bifurcation and wavenumber selection revisited. *J. Fluid Mech.* **777**, R2.

SÁNCHEZ, P. S., FERNÁNDEZ J. TINAO I. & PORTER, J. 2019 Vibroequilibria in microgravity: Comparison of experiments and theory. *Phys. Rev. E* **100**, 063103.

SÁNCHEZ, P. S., YASNOU V. GAPONENKO Y. MIALDUN A. PORTER J. & SHEVTSOVA, V. 2019 Interfacial phenomena in immiscible liquids subjected to vibrations in microgravity. *J. Fluid Mech.* **865**, 850–883.

TALIB, E., JALIKOP, S. V. & JUEL, A. 2007 The influence of viscosity on the frozen wave instability: theory and experiment. *J. Fluid Mech.* **584**, 45–68.

TINAO, I., PORTER, J., LAVERÓN-SIMAVILLA, A. & FERNÁNDEZ, J. 2014 Cross-waves excited by distributed forcing in the gravity-capillary regime. *Phys. Fluids* **26**.

TIPTON, C. R. & MULLIN, T. 2004 An experimental study of faraday waves formed on the interface between two immiscible liquids. *Phys. Fluids* **16**, 2336–2341.